# Kitaev's exact solution approximated

**For 25 years condensed matter physicists have searched for a material that realizes a macroscopic quantum state of matter, the quantum spin liquid. Recent experiments show that the necessary interaction may be found in a family of hexagonal ruthenium based materials.**

N. Peter Armitage

How do we classify states of matter?  In materials, in the most conventional cases, we classify their patterns of order by the symmetries they break. For instance, in a ferromagnet, all spins point a particular direction, a situation less symmetric than the high temperature state where spins point randomly. There are also states that break no symmetries, but assume patterns of order that are subtly encoded in their quantum mechanical wavefunctions in a fashion that still allows us to distinguish one state from another.  These *topological* states may possess a slew of interesting behavior including hosting unconventional particles that are fractions of conventional ones. Banerjee and colleagues report[1] experiments that identify the material $\alpha$ - RuCl$_3$ as perhaps the best known example of a system that hosts a specific anisotropic interaction that can drive a particular interesting topological state, the Kitaev quantum spin liquid (QSL).

In the conventional scenario when symmetries are broken at low temperatures, it is appropriate to refer to such states in the limit of zero temperature as "classical ground states". Even though the underlying interactions and microscopic degrees of freedom are indisputably quantum, the ground state is classical in the sense that it shows patterns of order that can be parameterized by a local observable, which frequently results in macroscopically detectable phenomena such as external fields.  A ferromagnet is the most obvious example of this phenomenon. Even for all the enduring discussion of superfluidity as a manifestation of "macroscopic quantum mechanics", a superfluid is actually a broken symmetry state where that most quantum mechanical of quantities – the wavefunction's phase – becomes a classical field.  The remarkable properties of a superfluid derive in large part from the fact that this field becomes a measurable classical parameter in much the same fashion as strain or magnetization is[3].

The notion of superposition and entanglement is central in quantum mechanics as applied to small systems, for example a particle can be both "here" and "there" simultaneously.  Given their "classical" nature, classical ground states can be generally expressed as an unentangled product of individual wavefunctions, which each express a local degree of freedom. In this regard conventional metals are far more quantum than superfluids as they are represented by a superposition of states where each electron has some probability to be in each of the single particle wavefunctions.  Topological phases, such as QSLs have assumed a prominent place in modern condensed matter physics, because they are different from classical states. In QSLs, due to frustration or competing interactions, classical order does not occur even down to the lowest temperatures. A superposition of states exists in which spins point in many different directions simultaneously giving true macroscopic quantum mechanics.  It is predicted that such states can possess exotic fractional excitations such as spin-½ particles that are each half of a normal spin-1 magnon excitation of a conventional ordered magnet.

The notion of a QSL goes back to Anderson's notion of a resonating valence bond in the 1970s, where he proposed it as the ground state of spins on a triangular lattice[5].  However, it has been challenging to demonstrate a given model has a QSL ground state as such ground states are usually inferred by approximations and inspired guesswork.   A remarkable exception is the 2006 theoretical tour de force of

Kitaev who constructed an exactly solvable model[6] of a QSL that considers spin-½ degrees of freedom on a 2D honeycomb lattice with a seemingly artificial anisotropic spin interaction, where only one of the different x,y,z spin components interact in each of the three bond directions (See Fig. 1).

In building up theoretical descriptions of many body systems, exact models are important despite their frequent artificiality because they establish the point-of-principle that a particular phase *could* exist. However, it was shown by Jackeli and Khaliullin[7,8] that even despite the contrived nature of the Kitaev model, its anisotropic interactions can actually manifest through the effects of strong spin-orbit coupling. These developments have set out an extensive search for materials that could be an explicit realization of the Kitaev model.  As strong spin orbit coupling is essential, iridates based on a hexagonal motif such as $Na_2IrO_3$ have been a natural place to look[9]. Unfortunately, neutron studies have been hampered by the large neutron absorption cross-section of Ir ions[10]. It also appears that most iridates have a substantial isotropic Heisenberg term[10], and despite the expected robustness of the Kitaev phase to such perturbations, most of these systems have classical ordered phases at low temperature.  Banerjee and co-workers report an interesting development in the search for the ideal Kitaev compound, in looking at the $\alpha$-$RuCl_3$ system[11].  Using inelastic neutron scattering they find a number of excitations with a temperature and frequency dependence consistent with a large Kitaev interaction.  $\alpha$-$RuCl_3$ also orders at low temperature, but its Heisenberg term is estimated to be smaller than in the iridates, making it perhaps the best example yet of a system with a dominating Kitaev interaction.  At low energies, the spectra appear to reflect the low temperature ordered state (the symmetry of which is that expected for an ordered state proximate to the Kitaev phase).  At higher energies the spectra more closely resemble that predicted for the Kitaev spin liquid.  This partitioning of energy scales is very similar to that found in magnetically ordered quasi-1D spin chain systems like $KCuF_3$, where the high energy behavior reflects that of the free fractionalized spin-1/2 particles of the spin liquid state inherent to 1D, but at low energies there is a crossover to spectra that reflect the ordered state[12].

Going forward it is important to find new materials or configurations that may reflect even closer idealities of exactly solvable models like the Kitaev model or systems that may surprise us in being realizations of topological QSLs.  It is also imperative for the community to find new ways of probing these systems that reveal their underlying non-trivial nature in a more explicit fashion.

*N. Peter Armitage is at the Institute for Quantum Matter, Department of Physics and Astronomy, The Johns Hopkins University, Baltimore, Maryland 21218, USA. e-mail:*
*npa@jhu.edu*

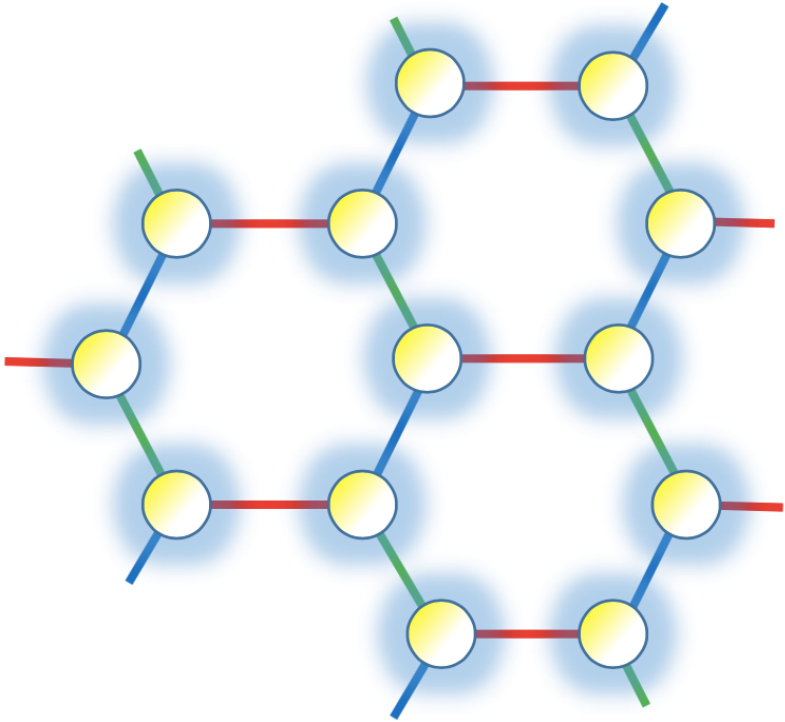

Figure 1: **Kitaev's quantum spin liquid**. Kitaev's model considers an exchange Hamiltonian on a hexagonal lattice with the form $\mathcal{H} = \sum_{i,j} K\, S_i^m S_j^m$, where $S_i^m$ and $S_j^m$ are the three different spin components ($m$ = $x$, $y$ or $z$) at two different nodes of the lattice ( denoted by $i$ and $j$) , and $K$ is the strength of the interaction between them. Along each of the three possible bond directions (here indicated by red, green, and blue lines) only a single spin component interacts with its neighbor (that is $S^x$ along red, $S^y$ along green, and is $S^z$ blue). In real materials it is expected that this interaction is supplemented by an isotropic Heisenberg term $\sum_{i,j} J\, \vec{S_i} \cdot \vec{S_j}$, where $J$ is the exchange constant and $\vec{S_i}$ and $\vec{S_j}$ the total spin at nodes $i$ and $j$.